\documentclass[11pt]{article}

\usepackage{amssymb}

\newtheorem{lem}{Lemma}

\newtheorem{theo}{Theorem}

\title{On the Approximation Method and the P versus NP Problem}

\author{Norbert Blum \\
        Institut f\"ur Informatik, Universit\"at Bonn \\    Endenicher Allee 19a,
D-53115 Bonn, Germany \\
email: blum@cs.uni-bonn.de}

\begin{document}

\maketitle


\begin{abstract}
  First of all we give some reasons that ``natural proofs'' built not a barrier to prove
  $\mbox{P} \not= \mbox{NP}$ using Boolean complexity. Then we investigate the approximation method for its
  extension to prove super-polynomial lower bounds for the non-monotone complexity of suitable Boolean
  functions in NP or to understand why this is not possible. It is given some evidence that the approximation
  method alone cannot be used to prove a super-linear lower bound for any function $f \in {\cal B}_n$.
  Additionally, an overview on the methods for proving lower bounds of the non-monotone and the monotone
  complexity of Boolean functions is given. Finally, a personal opinion how to proceed the research on the
  P versus NP problem and also on proving a super-linear lower bound for the non-monotone complexity of
  a Boolean function in NP is given.
\end{abstract}

\section{Introduction and Preliminaries}

Understanding the power of negations is one of the most challenging problems in complexity
theory. With respect to monotone Boolean functions, Razborov \cite{Ra3} was the first who could show
that the gain, if using negations, can be super-polynomial in comparision to monotone Boolean networks.
Tardos \cite{Ta} has improved this to exponential.
For the characteristic function of an NP-complete problem like the clique function,
it is widely believed that negations cannot help enough to improve the Boolean complexity from
exponential to polynomial. Since the computation of an one-tape Turing machine can be simulated
by a non-monotone Boolean network of size at most the square of the number of steps \cite[Chapter 3.9]{Sa},
a super-polynomial lower bound for the non-monotone network complexity of such a function would imply
$\mbox{P} \not= \mbox{NP}$. For the monotone complexity of such a function, exponential lower
bounds are known \cite{An,An2,Ra1,AlBo,Ka,Ha,Ju,BeUl,AmMa,HaRa}. But until now, no one could prove a super-linear
lower bound for the non-monotone complexity of any Boolean function in NP. An obvious attempt to
get a super-polynomial lower bound for the non-monotone complexity of the clique function could
be the extension of the method which has led to the proof of an exponential lower bound of its
monotone complexity. This is the so-called ``method of approximation'' developed in 1984 independently by
Andreev \cite{An} and Razborov \cite{Ra1}. In 1989, at the 21st STOC, Razborov \cite{Ra2} has presented the
sketch of a proof that his approximation method cannot be used to prove better than quadratic lower bounds for
the non-monotone complexity of any Boolean function.
But Razborov uses a very strong distance measure in his proof for the inability of the approximation
method. As elaborated in \cite{Bl1}, one can use the approximation method with a weaker distance measure
to prove super-polynomial lower bounds. 
Our goal is the extension of the approximation method to non-monotone Boolean networks to prove a super-polynomial
lower bound for the non-monotone complexity of a function in NP or to understand why this is not possible.

Firstly, we give some basic definitions. ${\cal B}_n := \{f \mid f:\{0,1\}^n \rightarrow \{0,1\}\}$ is
the set of all $n$-ary Boolean functions. The {\em $i$th variable\/} is denoted by 
$x_i:\{0,1\}^n \rightarrow \{0,1\}$, $1 \leq i \leq n$. Let $V_n := \{x_i \mid 1 \leq i \leq n\}$ and
$\overline{V}_n := \{\neg x_i \mid 1 \leq i \leq n\}$. Variables and negated variables are called
{\em literals\/}. A function $m:\{0,1\}^n \rightarrow \{0,1\}$ which is the conjunction of some 
literals is called a {\em monomial}. If we delete some literals from a monomial $m$ then we
obtain a {\em submonomial} $m'$ of $m$ and we write $m' \subseteq m$. The empty monomial $\varepsilon$ is the
constant function 1. The disjunction of monomials is a formula in {\em disjunctive normal form} (DNF).
The disjunction of some literals is called a {\em clause\/}. If we delete some literals from a
clause $d$ then we obtain a {\em subclause\/} of $d$. The empty clause is the constant
function 0. The conjunction of clauses is a formula in {\em conjunctive normal form} (CNF).
A monomial $m$ is called an {\em implicant\/} of the function $f$ if for all $a \in \{0,1\}^n$,
$m(a) = 1$ implies $f(a) = 1$. An implicant $m$ is a
{\em prime implicant\/} of $f$ if no proper submonomial of $m$ is an implicant of $f$. 
A clause $d$ is called an {\em $f$-clause\/} if for all $a \in \{0,1\}^n$, $d(a) = 0$ implies
$f(a) = 0$. A {\em prime clause\/} $d$ of $f$ is an $f$-clause where no proper subclause of $d$
is an $f$-clause. Let $a := (a_1,a_2, \ldots,a_n)$, $b := (b_1,b_2, \ldots,b_n) \in \{0,1\}^n$. 
We write $a \leq b$ iff $a_i \leq b_i$ for $1 \leq i \leq n$. A function $f \in {\cal B}_n$ is 
{\em monotone\/} iff $a \leq b$ implies $f(a) \leq f(b)$ for all $a,b \in \{0,1\}^n$.
Let $\Omega_0 := \{\wedge,\vee,\neg\}$ and $\Omega_m := \{\wedge,\vee\}$. For
$\Omega \in \{\Omega_0,\Omega_m, {\cal B}_2\}$, an {\em $\Omega$-network\/}
$\beta$ is a directed, acyclic graph such that each node has indegree at most two. The nodes $g$ with
indegree zero are {\em input nodes\/} and are labelled with $\mbox{op}(g) \in V_n \cup \{0,1\}$. The nodes $g$
with indegree larger than zero are the {\em gates\/} of $\beta$. Each gate $g$ is labelled with an operator
$\mbox{op}(g) \in \Omega$ where the indegree of $g$ is equal the number of operands of $\mbox{op}(g)$.
A node with outdegree zero is an {\em output node\/}.
For a node $g$ in $\beta$ let $\mbox{pred}(g) := \{h \mid h \rightarrow g \mbox{ is an edge in } \beta \}$ be
the set of its direct predecessors. With each node $g$, we associate a function
$\mbox{res}_{\beta}(g):\{0,1\}^n \rightarrow \{0,1\}$ which is defined by
$$
\mbox{res}_{\beta}(g) := \left\{ \begin{array}{ll}
                            \mbox{op}(g) & \mbox{$g$ is an input node,} \\
                 \neg \mbox{res}_{\beta}(h_1) & \mbox{op}(g) = \neg, \mbox{ } \mbox{pred}(g) = \{h_1\}, \\
   \mbox{res}_{\beta}(h_1) \: \mbox{op}(g) \: \mbox{res}_{\beta}(h_2) & \mbox{otherwise, where } 
                   \mbox{pred}(g) = \{h_1,h_2\}. \\
                            \end{array}
\right.
$$

The functions $\mbox{res}_{\beta}(g)$ with $g$ is a node in $\beta$ are computed by $\beta$.
Let $f \in {\cal B}_n$. The minimum number of gates in an $\Omega_0$-network which computes $f$ where 
negations are not counted is the {\em non-monotone complexity\/} $C(f)$ of $f$. Each operator
in ${\cal B}_2$ can be realized by an $\Omega_0$-network using at most two gates. Hence, for proving a
super-linear lower bound for the size of a ${\cal B}_2$-network realizing a Boolean function $f$ we can restrict
us to prove a super-linear lower bound for the non-monotone complexity of $f$.
An $\Omega_m$-network is called a {\em monotone network}. Note that exactly the monotone Boolean functions
can be computed by a monotone network. The minimum number of gates in a 
monotone network which computes the monotone function $f$ is the {\em monotone complexity\/} 
$C_m(f)$ of $f$. 

Given any $\Omega_0$-network $\beta$, we can convert $\beta$ to an equivalent 
$\Omega_0$-network $\beta'$ where all negations occur only at the input nodes. Moreover,
the size of $\beta$ is at most doubled. For doing this, we start at the output 
nodes and apply De Morgan rules for bringing the negations to the input nodes. Since gates can
be simultaneously negated and non-negated, some gates have to be doubled. 
The resulting network is a so-called {\em standard network\/} where only input variables are negated.
We consider a negated variable $\neg x_i$ as an input node $g$ with $\mbox{op}(g) = \neg x_i$.
The {\em standard complexity\/} 
$C_{st}(f)$ of a function $f \in {\cal B}_n$ is the size of a smallest standard network which computes $f$.
Note that the standard and the non-monotone complexity of a function $f$ differs at most by the
factor two. Hence, for proving a super-linear lower bound for the non-monotone complexity of
a Boolean function, we can restrict us to the consideration of standard networks.

Before the investigation of the approximation method with respect to its extension to standard networks, we have
to deal with the paper of Razborov and Rudich \cite{RaRu}.
In 1994, Razborov and Rudich have introduced the notion of a ``natural proof''. They say that the known
proofs of lower bounds on the complexity of explicit Boolean functions in non-monotone models fall within their
definition of natural. They have shown that natural proofs cannot be used for separating P and NP unless hard
pseudorandom generators do not exist. Since the existence of such generators is widely believed, natural proofs 
are widely accepted to be a barrier for proving $\mbox{P} \not= \mbox{NP}$ using Boolean complexity. We discuss
this in the subsequence.

Firstly, we give their definition of a ``natural proof''. This is a proof which uses a natural combinatorial
property. A {\em combinatorial property} is a subset $\{C_n \subset {\cal B}_n \mid n \in \mathbb{N}\}$ of
Boolean functions. $C_n$ is called {\em natural} if there is $C^*_n \subseteq C_n$ which satisfies:
\begin{enumerate}
\item
  For all $f \in {\cal B}_n$ it can be decided in $2^{O(n)}$ time if $f \in C^*_n$. ({\em constructiveness})
\item
  $|C^*_n| \geq 2^{-O(n)}|{\cal B}_n|$. ({\em largeness})
\end{enumerate}
The first property means that the characteristic function of $C^*_n$ can be computed in polynomial time in the
size of the truth table of the input function $f \in {\cal B}_n$. The second property says that a function
randomly chosen from ${\cal B}_n$ is contained in $C^*_n$ with non-negligible probability. $\mbox{P}/\mbox{poly}$
is the set of languages which are recognizable by a family of Boolean networks of polynomial size.
Note that $\mbox{P} \subseteq \mbox{P}/\mbox{poly}$. A combinatorial property is
{\em useful against $\mbox{P}/\mbox{poly}$} if the network complexity of any sequence
$f_1,f_2, \ldots, f_n, \ldots$ where $f_n \in C_n$ is super-polynomial; i.e., for all $k \in \mathbb{N}$ there is
$n_k \in \mathbb{N}$ such that the network complexity of $f_n$ is larger than $n^k$ for all $n > n_k$. A proof
that a Boolean function does not have polynomial network complexity is
{\em natural against $\mbox{P}/\mbox{poly}$} if the proof uses a natural combinatorial property $C_n$ which is
useful against $\mbox{P}/\mbox{poly}$.

Razborov and Rudich mention that ``from experience it is plausible to say that we do not yet understand the
mathematics of $C_n$ outside exponential time (as a function of $n$) well enough to use them effectively in a
combinatorial style proof.'' This means that combinatorial properties used in a today's lower bound proof are
constuctive.
With respect to the largeness property, they write: ``In Section 5 we give some solid theoretical evidence for
largeness, by showing that any $C_n$ based on a {\em formal complexity measure} must be large.''  We
discuss now the ``solid theoretical evidence for largeness'' given by Razborov and Rudich.

A {\em formal complexity measure} is a function $\mu:{\cal B}_n \mapsto \mathbb{R}^+$ such that
\begin{itemize}
\item[a)]
  $\mu(f) \leq 1$ for $f \in \{x_1,x_2, \ldots,x_n,\neg x_1,\neg x_2, \ldots,\neg x_n\}$, and
\item[b)]
  $\mu(f \wedge g) \leq \mu(f) + \mu(g)$ and
  $\mu(f \vee g) \leq \mu(f) + \mu(g)$ for all $f,g \in {\cal B}_n$.
\end{itemize}

A {\em formula} is a Boolean network where the underlying graph is a tree. The {\em size} of a formula $\beta$
is the number of leaves in $\beta$. The {\em formula size} $L_{\Omega_0}(f)$ of a Boolean function
$f \in {\cal B}_n$ is the size of a smallest $\Omega_0$-formula which computes $f$. Note that $L_{\Omega_0}$
itself is a formal complexity measure. Moreover, by induction on the formula size, it can be shown that
$L_{\Omega_0}(f) \geq \mu(f)$ for all $f \in {\cal B}_n$ and each formal complexity measure $\mu$.

Razborov and Rudich show that ``any formal complexity measure $\mu$ which takes a large value at a single function,
must take large values almost everywhere.'' This is formalized by the following theorem.
\begin{theo} \label{theo1.1}
  Let $\mu$ be a formal complexity measure on ${\cal B}_n$, and let $\mu(f) \geq t$ for some $f \in {\cal B}_n$.
  Then for at least $1/4$ of all functions $g \in {\cal B}_n$, $\mu(g) \geq t/4$.
\end{theo}
Razborov and Rudich conclude that ``every combinatorial property based on such a measure automatically satisfies
the largeness condition in the definition of natural property.'' But they do not formalize what they mean that a
property is based on a formal complexity measure. The combinatorial property of $f \in {\cal B}_n$ cannot be
``$\mu(f) \geq t$'' for a certain bound $t$ since the property is used to prove this lower bound. Let
$C_n \subset {\cal B}_n$ be the combinatorial property used to prove $\mu(f) \geq t$; i.e., $f \in C_n$. 
Theorem \ref{theo1.1} does not imply that a function $g$ having the same measure as $f$ has to be contained in
$C_n$. Hence, Theorem \ref{theo1.1} does not imply the largeness property for $C_n$.

Theorem \ref{theo1.1} is not a surprise because of Shannon's famous counting argument \cite{Sh,We} which shows
that at least a fraction of $(1 - 2^{-2^nn^{-1} \log\log n})$ of the functions in ${\cal B}_n$ has non-monotone
complexity at least $\frac{2^n}{n}$. The theorem tells us that we can only use a formal complexity measure to
prove a lower bound $t$ for the non-monotone network complexity of a function $f \in {\cal B}_n$ which has the
property that up to the constant factor $\frac{1}{4}$, the measure of at least a quarter of the functions in
${\cal B}_n$ is at least as large as the measure of $f$. Using the counting argument again, we know that the
formula size has this property. By Shannon's counting argument, a combinatorial property which implies that a
function having this property has complexity smaller than $\frac{2^n}{n}$ could never fulfill the largeness
property. Hence, a proof which uses such a combinatorial property is not natural.

Altogether, it seems that ``natural proofs'' built not a barrier for proving $\mbox{P} \not= \mbox{NP}$ using
Boolean complexity. Therefore, it makes sence to investigate the approximation method with regard its expandability
to standard networks. This is the aim of the paper.

The first problem which we have to investigate is the treatment of the negated variables.
In \cite{Bl2}, I have tried to treat the negated variables in a standard network which computes a given
non-constant monotone Boolean function $f$ at its output node $g_t$ in such a way that we can use the
approximators developed for $f$ with respect to monotone networks on standard networks. The proof of Theorem 6
in \cite{Bl2} is wrong. The mistake in the proof is explained in \cite{Bl3}. The motivation of the approach in
\cite{Bl2} was the avoidance of the explicit consideration of the negated variables. The conclusion in \cite{Bl3}
is that the negated variables have to be approximated as well. Therefore, we have to consider the negated
variables explicitely. Before doing this, a lot of work has to be done.

We shall investigate the computation in standard networks in the next section. In Section 3, an overview on the
methods for proving lower bounds of the non-monotone and of the monotone complexity of Boolean functions is given.
We describe the general idea of the approximation method in Section 4. For monotone networks, two kinds of
approximators are known, the CNF-DNF-approximators and the sunflower-approximators. Section 5 is devoted the
description of these approximators with respect to monotone networks. The extension of these approximators to
standard networks is treated in Section 6. It is given evidence that CNF-DNF- and also sunflower-approximators
alone cannot be used to prove a super-linear lower bound for the standard complexity of any function in
${\cal B}_n$. In Section 7, a personal opinion how to proceed the research on the P versus NP problem and also on
proving a super-linear lower bound for the non-monotone complexuty of a Boolean function in NP is given.

\section{The Computation in Standard Networks}

Let $\beta$ be a standard network which computes a function $f \in {\cal B}_n$ at its output node $g_t$. Let $g$
be any node in $\beta$. The function $\mbox{res}_{\beta}(g)$ can be written as a DNF-formula; i.e.,
$\mbox{res}_{\beta}(g) = \bigvee_{j=1}^r m_j$ where each $m_j$ is a monomial. We call this representation of
$\mbox{res}_{\beta}(g)$ the {\em DNF-representation\/} $\mbox{DNF}_{\beta}(g)$ of $\mbox{res}_{\beta}(g)$. The function
$\mbox{res}_{\beta}(g)$ can be written as a CNF-formula as well; i.e., $\mbox{res}_{\beta}(g) = \bigwedge_{j=1}^s d_j$
where each $d_j$ is a clause. We denote this formula the {\em CNF-representation\/} $\mbox{CNF}_{\beta}(g)$ of
$\mbox{res}_{\beta}(g)$.
In contrast to monotone networks, $\mbox{DNF}_{\beta}(g_t)$ must not contain the prime implicants of the function
$f$ as monomials. Furthermore, $\mbox{CNF}_{\beta}(g_t)$ must not contain the prime clauses of the function $f$ as
clauses. To extend the proof techniques developed for monotone networks to standard networks, it is useful to
recognize the prime implicants in $\mbox{DNF}_{\beta}(g_t)$.
Similarly, it is useful to recognize the prime clauses in $\mbox{CNF}_{\beta}(g_t)$. For doing
this, let $p_1,p_2, \ldots,p_k$ be the prime implicants and $c_1,c_2, \ldots,c_s$ be the prime clauses of the
function $f$. Each monomial $m_j$ in  $\mbox{DNF}_{\beta}(g_t) = \bigvee_{j=1}^{t_o} m_j$ is an implicant of the
function $f$. Otherwise, it would exist an input $(a_1,a_2, \ldots,a_n) \in \{0,1\}^n$ such that
$f(a_1,a_2, \ldots,a_n)  = 0$ but $\mbox{res}_{\beta}(g_t)(a_1,a_2, \ldots,a_n) = 1$. This means that each monomial in
$\mbox{DNF}_{\beta}(g_t)$ contains at least one prime implicant of the function $f$ as a submonomial.
If a monomial $m_j$ contains $l > 1$ prime implicants then we add $l-1$ copies of $m_j$ to
the DNF-representation. By separating in each monomial containing a prime implicant $p_j$ the prime
implicant and the other literals, we can write 
$$\mbox{res}_{\beta}(g_t) = \bigvee_{j=1}^k \bigvee_{i=1}^{l_j} p_j \wedge m'_{j_i},$$
where $\mbox{DNF}_{\beta}(g_t)$ contains $l_j$ monomials including the prime implicant $p_j$, $1 \leq j \leq k$. 

Similarly, we separate in each $f$-clause $d_j$ in $\mbox{CNF}_{\beta}(g_t)$ the contained prime clause from
the other literals. If a clause $d_j$ contains $l > 1$ prime clauses then we add $l-1$ copies of $d_j$ to
the CNF-representation. By separating in each clause containing a prime clause $c_j$ the prime
clause and the other literals, we can write 
$$\mbox{res}_{\beta}(g_t) = \bigwedge_{j=1}^s \bigwedge_{i=1}^{l_j} c_j \vee d'_{j_i},$$
where $\mbox{CNF}_{\beta}(g_t)$ contains $l_j$ clauses including the prime clause $c_j$, $1 \leq j \leq s$.

Now we describe the DNF- and CNF-formulas constructed by a standard network $\beta$.
Note that after a simplification of the network, no input node $g$ with $\mbox{op}(g) \in \{0,1\}$ exists.
Assume that for all input nodes $g$, $op(g) \in V_n \cup \overline{V}_n$.
Starting at the input nodes, the network $\beta$ constructs the DNF-formulas in the following way:
\begin{enumerate}
\item
  If $g$ is an input node with $\mbox{op}(g) = x_i$ or $\mbox{op}(g) = \neg x_i$ then 
  $$\mbox{DNF}_{\beta}(g) := \mbox{op}(g).$$
\item
  If $g$ is an $\vee$-gate with $\mbox{pred}(g) = \{h_1,h_2\}$ then 
  $$\mbox{DNF}_{\beta}(g) := \mbox{DNF}_{\beta}(h_1) \vee \mbox{DNF}_{\beta}(h_2).$$
\item
  If $g$ is an $\wedge$-gate with $\mbox{pred}(g) = \{h_1,h_2\}$, $\mbox{DNF}_{\beta}(h_1) = \bigvee_{i=1}^{t_1} m_i$
  and $\mbox{DNF}_{\beta}(h_2) = \bigvee_{j=1}^{t_2} m'_j$ then
  $$\mbox{DNF}_{\beta}(g) := \bigvee_{i=1}^{t_1}\bigvee_{j=1}^{t_2} (m_i \wedge m'_j).$$
\end{enumerate}

Each input $a \in \mbox{res}_{\beta}(g)^{-1}(1)$ satisfies a monomial $m_j$ of $\mbox{DNF}_{\beta}(g)$.
Each input $b \in \mbox{res}_{\beta}(g)^{-1}(0)$ does not satisfy any monomial in $\mbox{DNF}_{\beta}(g)$. Hence,
each monomial in $\mbox{DNF}_{\beta}(g)$ contains a variable $x_i$ with $b_i = 0$ or a negated variable
$\neg x_j$ with $b_j = 1$.

Starting at the input nodes, the network $\beta$ constructs the CNF-formulas in the following way:
\begin{enumerate}
\item
  If $g$ is an input node with $\mbox{op}(g) = x_i$ or $\mbox{op}(g) = \neg x_i$ then 
  $$\mbox{CNF}_{\beta}(g) := \mbox{op}(g).$$
\item
  If $g$ is an $\wedge$-gate with $\mbox{pred}(g) = \{h_1,h_2\}$ then 
  $$\mbox{CNF}_{\beta}(g) := \mbox{CNF}_{\beta}(h_1) \wedge \mbox{CNF}_{\beta}(h_2).$$
\item
  If $g$ is an $\vee$-gate with $\mbox{pred}(g) = \{h_1,h_2\}$, $\mbox{CNF}_{\beta}(h_1) = \bigwedge_{i=1}^{t_1} d_i$
  and $\mbox{CNF}_{\beta}(h_2) = \bigwedge_{j=1}^{t_2} d'_j$ then
  $$\mbox{CNF}_{\beta}(g) := \bigwedge_{i=1}^{t_1}\bigwedge_{j=1}^{t_2} (d_i \vee d'_j).$$
\end{enumerate}

Each input $b \in \mbox{res}_{\beta}(g)^{-1}(0)$ falsifies a clause $d_j$ of $\mbox{CNF}_{\beta}(g)$.
Each input $a \in \mbox{res}_{\beta}(g)^{-1}(1)$ does not falsify any clause in $\mbox{CNF}_{\beta}(g)$. Hence, each
clause in $\mbox{CNF}_{\beta}(g)$ contains a variable $x_i$ with $a_i = 1$ or a negated variable $\neg x_j$
with $a_j = 0$.

The following theorem characterizes exactly the DNF-representation and the CNF-representation of
$\mbox{res}_{\beta}(g_t)$ with respect to a standard network which computes a Boolean function $f \in {\cal B}_n$ at
its output node $g_t$. 
\begin{theo} \label{theo2.1}
  Let $\beta$ be a standard network which computes a Boolean function $f \in {\cal B}_n$ at its
  output node $g_t$. Then the following hold:
  \begin{itemize}
  \item[a)]
    $\mbox{DNF}_{\beta}(g_t)$ contains only implicants of the function $f$. Furthermore,
    for each $a \in f^{-1}(1)$, $\mbox{DNF}_{\beta}(g_t)$ contains an implicant $m_a$ of $f$ such that $m_a(a) = 1$.
  \item[b)]
    $\mbox{CNF}_{\beta}(g_t)$ contains only $f$-clauses. Furthermore, for each
    $b \in f^{-1}(0)$, $\mbox{CNF}_{\beta}(g_t)$ contains an $f$-clause $d_b$ such that $d_b(b) = 0$.
  \end{itemize}
\end{theo}
{\bf Proof}:
Assume that $\mbox{DNF}_{\beta}(g_t)$ contains a monomial $m$ which is not an implicant of $f$. Then,
by the definition of an implicant of $f$, there exists $b \in \{0,1\}^n$ such that $m(b) = 1$ but $f(b) = 0$.
This contradicts the assumption that $\beta$ computes $f$ at its output node $g_t$. Hence, all monomials
of $\mbox{DNF}_{\beta}(g_t)$ are implicants of $f$.

Assume that there is $a \in f^{-1}(1)$ such that $m(a) = 0$ for all implicants $m$ in $\mbox{DNF}_{\beta}(g_t)$. Then
$\mbox{res}_{\beta}(g_t)(a) = 0$ but $f(a) = 1$. This contradicts the assumption that $\beta$ computes $f$ at its
output node $g_t$. Hence, for each $a \in f^{-1}(1)$, $\mbox{DNF}_{\beta}(g_t)$ contains an implicant $m_a$ of $f$
such that $m_a(a) = 1$.

This proves part a) of the theorem. Analogously, part b) of the theorem can be proved.
$\Box$

\smallskip
Every DNF-formula can be transformed into an equivalent CNF-formula. To see this let
$\alpha = \bigvee_{i=1}^{t_0} m_i$ be a DNF-formula which computes a Boolean function $f \in {\cal B}_n$. To obtain an
equivalent CNF-formula $\gamma$, we pick from each monomial $m_i$, $1 \leq i \leq t_0$ one literal and perform the
disjunction of all chosen literals. Then the conjunction of all clauses which can be constructed in this way
is a CNF-formula $\gamma = \bigwedge_{j=1}^{s_0} d_j$ which corresponds to the DNF-formula $\alpha$. The following lemma
shows that $\gamma$ computes the function $f$.
\begin{lem}  \label{lem2.1}
  Let $\alpha = \bigvee_{i=1}^{t_0} m_i$ be a DNF-formula which computes a Boolean function $f \in {\cal B}_n$. Let
  $\gamma = \bigwedge_{j=1}^{s_0} d_j$ be the CNF-formula constructed from $\alpha$ as described above. Then $\gamma$
  computes $f$.
\end{lem}
{\bf Proof}:
Consider $a \in f^{-1}(1)$. Then there is a monomial $m_l$ in $\alpha$ such that $m_l(a) = 1$. Since each clause of
$\gamma$ contains a literal of $m_l$, the input $a$ satisfies all clauses in $\gamma$. Hence $\gamma(a) = 1$.

Let $b \in f^{-1}(0)$. Then each monomial in $\alpha$ contains a literal which is not satisfied by $b$. Consider a
clause $d_l$ of $\gamma$ which picks from each monomial a literal which is not satisfied by $b$. Obviously,
$d_l(b) = 0$. Hence, $\gamma(b) = 0$.

Altogether, we have shown that $\gamma$ computes $f$.
$\Box$

\smallskip
We call such a transformation of a DNF-formula to an equivalent CNF-formula a {\em DNF/CNF-switch\/}.
A DNF/CNF-switch can be organized as the construction of a tree $T$ in the following way:
\begin{enumerate}
\item
  Each edge in $T$ is labelled by a literal. With each node $w$ in $T$ we associate the clause $d(w)$ which is
  obtained by the disjunction of the literals on the unique path from the root of $T$ to $w$. $T$ is constructed
  while {\em expanding\/} the monomials $m_0,m_1,m_2, \ldots,m_{t_0}$ where $m_0$ is the empty monomial.
\item
  While expanding $m_0$, the root of $T$ is created. The associated clause is the empty clause.
\item
  Suppose that $w$ is a leaf that was created while expanding $m_i$. Then the monomial $m_{i+1}$ is expanded at
  the leaf $w$ in the following way: The leaf $w$ obtains for each literal in $m_{i+1}$ a new son $w'$. The edge
  $(w,w')$ is labelled with the corresponding literal.
\end{enumerate}
After the construction of the tree $T$, the clauses corresponding to the paths from the root of $T$ to the
leaves are the clauses contained in the CNF-formula $\gamma$ obtained from $\alpha = \bigvee_{i=1}^{t_0} m_i$ by
performing a DNF/CNF-switch.

Analogously, every CNF-formula can be transformed into an equivalent DNF-formula.
We call such a transformation of a CNF-formula to an equivalent DNF-formula a {\em CNF/DNF-switch\/}.


\section{On Proof Methods in Boolean Complexity}

To get a lower bound for the size of a network $\beta$ which computes a Boolean function $f \in {\cal B}_n$, we
have to count gates in $\beta$. The problem is that we have no knowledge about the structure of $\beta$; i.e.,
we can only use the fact that the network $\beta$ computes the function $f$. Therefore, with respect to a complete
basis ${\cal B}_2$ or $\Omega_0$, only small linear lower bounds for the network complexity of a function in NP
could be proved. If the function $f$ depends on each of the $n$ input variables then each of the $n$ input nodes
has to be connected to the output node. For doing this, at least $n-1$ gates with indegree two are needed.
Hence, each Boolean network which computes a function depending on all $n$ input variables contains at least $n-1$
gates. Since functions like $x_1 \wedge x_2 \wedge \ldots \wedge x_n$ depend on all variables and can be realized
with only $n-1$ gates, without an additional argument, no larger lower bound can be proved. Slightly better lower
bounds are
obtained using the so-called {\em gate-elimination method\/}. The gate-elimination method uses induction. By
an assignment of some variables with values from $\{0,1\}$, a specific small constant number of gates is
eliminated in each step and the resulting function is of the same type as the function before the assignment. 
Over the years, the case analyses used in the proofs have become more and more complicated impoving the lower
bounds only slightly. For an overview see \cite{We,Bl1,FiGoHiKu}. I am convinced that the elimination method alone
cannot be used to prove a super-linear lower bound for the network complexity of any function in NP.

What happen if we consider Boolean functions with many outputs as the Boolean matrix multiplication or the
Boolean convolution? With respect to the convolution, each output depends on nearly all variables. Moreover,
as shown by Valiant \cite{Va}, the graph of any network computing the convolution is an $n$-superconcentrator.
An {\em $n$-superconcentrator\/} is a directed graph with $n$ input and $n$ output nodes such that for each
subset of the input nodes and each subset of the output nodes of the same size $r$ there are $r$ mutually
node-disjoint paths connecting the set of input nodes with the set of output nodes. Aho, Hopcroft and Ullman
\cite{AhHoUl} have conjectured that an $n$-superconcentrator has at least $n \log n$ edges. But Valiant
\cite{Va} itself has shown that there exist superconcentrators of linear size destroying the hope to prove
a super-linear lower bound using graph theoretical arguments only. Also for functions with many outputs, no
super-linear lower bound for its network complexity is known.

The inability to prove super-linear lower bounds for the non-monotone complexity of explicit Boolean functions
has led to the consideration of restricted models of Boolean networks like monotone or bounded-depth Boolean
networks. For both restricted models, exponential lower bounds for the complexity of an explicit Boolean
function in NP are known. We are interested in proving a super-linear lower bound of the non-monotone complexity 
of a Boolean function in NP. Bounding the depth of the network to be constant seems to be a much harder restriction 
than allowing only monotone networks. Some functions with linear network complexity are used to prove an exponential
lower bound for the size of a constant depth network computing the function. Techniques for proving super-linear
lower bounds for the monotone complexity of functions which are also candidates for proving a super-linear lower
bound for its non-monotone complexty seems to be more suitable for their extension to get a super-linear lower
bound for the non-monotone complexity of a function in NP. Therefore, we are interested in the methods developed
for the proof of lower bounds for the monotone complexity of Boolean functions.

The core of each super-linear lower bound proof for the monotone complexity of a Boolean function is the
successful application of certain replacement rules. In a monotone network $\beta$ computing a Boolean
function $f$, a {\em replacement rule\/} replaces a node $u$ with $\mbox{res}_{\beta}(u) = h$ by a node $u'$ which
computes a function $h'$. In most cases, $h'$ depends on $h$. The first replacement rules used to prove some
lower bounds have the additional property that the resulting monotone network $\beta'$ still computes the function
$f$. Such a replacement rule is of {\em Type 1\/}. For a presentation of replacement rules of Type 1 see
\cite{MeGa, We}.

Replacement rules of Type 1 in combination with the gate elimination method are used to prove
lower bounds for the monotone complexity of Boolean sums \cite{Ne,Pi,We2,Me}, Boolean matrix multiplication
\cite{Pr,Pa,MeGa} and generalized Boolean matrix multiplication \cite{We1}. Replacement rules of Type 1
are used in different ways. With respect to Boolean matrix multiplication \cite{Pa,MeGa}, they are used
for the characterization of an optimal monotone network. For the generalized matrix product \cite{We1} and
for Boolean sums \cite{We2,Me} they are used explicitely; i.e., the gate $u$ is replaced by a subnetwork which
computes the function $h'$. This should be possible without additional cost. To get this, Wegener \cite{We1} has
introduced a technique, very common in algebraic complexity, into Boolean complexity. Certain functions are given
for free as inputs of the network. A lower bound for such a network implies the same lower bound for the monotone
complexity of the considered function. Wegener \cite{We1} uses this technique in combination with the gate
elimination method to prove an $\Omega(n^2/\log^2 n)$ lower bound for the monotone complexity of the generalized
Boolean matrix product. In \cite{We3}, Wegner has introduced a further technique improving this lower bound
to $\Omega(n^2/\log n)$. Instead using the gate elimination method, he has defined a suitable {\em value function\/}
to estimate the contribution of each $\wedge$-gate for the computation of the outputs. At each $\wedge$-gate,
the value function distributes at most the value $1$ among the prime implicants. Then he has proved the necessity
to give to each prime implicant at least the value $\frac{1}{2}$ obtaining a lower bound of half the number of
prime implicants. The definition of the value function depends on the structure of the function computed at the
$\wedge$-gate under consideration. Important for the proof is that the function has many outputs and also the structure
of the prime implicants of the functions computed at the output nodes.

The first super-linear lower bound for the monotone complexity of an explicit Boolean function has been proved by
Neciporuk \cite{Ne} in 1969 for a function in ${\cal M}_{n,n}$, a set of so-called Boolean sums. A {\em Boolean sum\/}
$f_i$ is the disjunction of a subset $F_i \subseteq V_n$ of variables. He has considered the monotone complexity of
sets of Boolean sums which have ``nothing in common''. Nothing in common means that two distinct Boolean sums have at
most one variable in common. We say then that the set of Boolean sums is {\em $(1,1)$-disjoint\/}. One can think that
$\wedge$-gates cannot reduce the monotone complexity of a set of Boolean sums in comparision to networks which use only
$\vee$-gates. But Tarjan \cite{Ta2,We} has given an example which shows that using $\wedge$-gates can reduce the monotone
complexity. For $(1,1)$-disjoint sets of Boolean sums, Neciporuk has proved that optimal monotone Boolean networks contain
only $\vee$-gates. A well known construction of K\H{o}v\'{a}ri, S\'{o}s and Tur\'an \cite{KoSoTu} leads to an explicitely
constructed $(1,1)$-disjoint set $f = (f_1,f_2,\ldots,f_n)$ of Boolean sums with $\Omega(n^{3/2})$ prime implicants such
that an $\Omega(n^{3/2})$ lower
bound for the monotone complexity of this function has been proved. Some years later, Pippenger \cite{Pi} and
Mehlhorn \cite{Me} have generalized the approach of Neciporuk to sets of Boolean sums which are {\em $(h,k)$-disjoint};
i.e., any $h+1$ different Boolean sums have at most $k$ variables in common. Such a set of Boolean sums corresponds to
a bipartite graph which does not contain a $K_{h+1,k+1}$ as a subgraph where $K_{h+1,k+1}$ denotes the complete bipartite
graph with node sets of sizes $h+1$ and $k+1$, respectively. Bipartite graphs can be represented by Boolean matrices. A
Boolean matrix $A$ is {\em $(h,k)$-free\/} if it does not contain any $(h+1) \times (k+1)$ submatrix containing only ones.
A bipartite graph contains no $K_{h+1,k+1}$ iff the corresponding Boolean matrix is $(h,k)$-free. The question about the
maximal number of ones in a $(h,k)$-free $(n \times n)$-matrix is the famous problem of Zarankievicz. Pippenger and
Mehlhorn showed that using $\wedge$-gates for the computation of a set of $(h,k)$-disjoint Boolean sums can save at most
the factor $\max\{h-1,k-1\}$. Using a construction of Brown \cite{Br}, a (2,2)-disjoint set of Boolean sums with
$\Omega(n^{5/3})$ prime implicants has been constructed such that an $\Omega(n^{5/3})$ lower bound for the monotone
complexity of this Boolean function has been proved. In the subsequence, the explicit construction of further dense
$(h,k)$-free Boolean matrices \cite{An1,KoRoSz} has led to larger lower bounds.

Boolean sums, the Boolean matrix multiplication and the generalized Boolean matrix multiplication have some
disjointness properties which the convolution does not have. Therefore, to prove a lower bound for the convolution,
the situation becomes more difficult. The first approach for proving a lower bound for the monotone complexity
of the convolution uses graph-theoretical properties of monotone networks realizing the convolution. Pippenger
and Valiant \cite{PiVa} have studied shifting graphs and have proved that each monotone network for the convolution
has to be a shifting graph obtaining an $\Omega(n \log n)$ lower bound for monotone complexity of the convolution.
To prove a lower bound of size $\Omega(n^{4/3})$ for the number of
$\wedge$-gates needed in a monotone network which computes the convolution, the author \cite{Bl0} has introduced
two further techniques into Boolean complexity. For the first time, a replacement rule changing the function computed
at the output nodes of the network is used. An $\wedge$-gate $g$ such that the function computed at the output of the gate
$g$ has a certain property is replaced by $0$. Therefore, the gate $g$ is eliminated but, at the output nodes, the
construction of some prime implicants could be destroyed. Because of the property of the function computed at the output
of the gate $g$, the number of destroyed prime implicants is bounded. Only for inputs such that an output has to be one,
a wrong value could be computed at the output node because of an application of the replacement rule. Such a replacement
rule is of {\em Type 2\/}. To apply the replacement rule, we need that the monotone network has a certain structure. With
respect to a given proof technique, we call a monotone network which allow the application of the proof technique a
{\em normal form network\/}. Given any monotone network $\beta$ computing a given monotone Boolean function, the
network is transformed into normal form first and then, the corresponding proof technique is applied. Maybe, the 
transformation increases the size of the network such that this increase has to be taken into consideration to obtain a
lower bound for the monotone complexity from the lower bound for a normal form network.
Wei{\ss} \cite{Wei} observed that on each path from the input node $a_i$ to an output node $c_k$ which depends on $a_i$ and
has at least two prime implicants there has to be a first $\vee$-gate such that some $a_jb_l$ with $j \not= i$ is an
implicant of the function. Using a replacement rule of Type 1, he showed that all these $\vee$-gates can be eliminated
after setting $a_i$ to zero. Consider the assignment $\alpha$ which we obtain after setting with respect to each such an
$\vee$-gate the variables of the prime implicant $a_jb_l$ to one. Then all outputs of the network do not depend on $a_i$.
Therefore for each output function $c_k$ which depends on $a_i$, the assignment $\alpha$ has to satisfy any prime
implicant of $c_k$. Since the Boolean convolution is semi-disjoint, each conjunction of a variable in
$A = \{a_0,a_1, \ldots,a_{n-1}\}$ and a variable in $B = \{b_0,b_1, \ldots,b_{n-1}\}$ is prime implicant of exactly one
output function. Therefore, at most $p^2$ prime implicants can be constructed if the assignment $\alpha$ is defined
with respect to $p$ first $\vee$-gates having the needed property. Since $n$ output functions depend on $a_i$, $p^2$
has to be at least $n$ such that an $n^{3/2}$ lower bound for the number of $\vee$-gates needed in a monotone network
which computes the convolution could be proved. Grinschuk and Sergeev \cite{GrSe} have constructed $(h,k)$-disjoint
Boolean circulant matrices with many ones. The complexity of the corresponding set of Boolean sums is
$\Omega(n^2 \log^{-6}n)$. Since circulant matrices are related to cyclic convolution and cyclic Boolean convolution
can be reduced to Boolean convolution \cite{GrSe,Ju3}, they obtain an $\Omega(n^2 \log^{-6}n)$ lower bound for the
number of $\vee$-gates in a monotone network computing the Boolean convolution. Therefore, the used proof technique
to prove the lower bound for the Boolean convolution was reduction.

Although since 1969 super-linear lower bounds for the monotone complexity of explicit functions in ${\cal M}_{n,m}$
where $m = \Theta(n)$ have been proved, before 1984, the largest lower bound for the monotone complexity of an explicit
single output function was of size $4n$ \cite{Ti}. All super-linear lower bound proofs for the monotone complexity of
functions in ${\cal M}_{n,m}$ strongly depend on the property that a set of functions has to be computed. With respect
to single output monotone Boolean functions, no technique for counting a super-linear number of gates has been developed
before 1984. In 1984, Andreev \cite{An,An2} and Razborov \cite{Ra1,Ra3} independently achieved the breakthrough.
They have proved super-polynomial lower bounds for certain single output functions in NP. The functions
$\mbox{res}_{\beta}(g)$ computed at the gates $g$ are replaced by a function which approximates $\mbox{res}_{\beta}(g)$.
The main point was the introduction of replacement rules which change the value of the function computed at the output
node with respect to inputs in $f^{-1}(0)$ where $f$ is the considered function. Such a replacement rule is of
{\em Type 3\/}. The so-called {\em approximation method\/} was born. In the next section, we will describe the
approximation method in detail. 
  
\section{The Approximation Method}

Both, Andreev and Razborov have used set theoretical constructions to prove the lower bound. In a sence, this hides the
effect of the approximation on the computation in the network. To understand this effect, we describe the approximation
method for monotone networks directly on a monotone network which computes the function under consideration. For the
extension of the approximation method to standard networks, this approach is more suitable than using a set theoretical
construction as Razborov and Andreev.

To get a lower bound for the monotone complexity of a monotone function $f \in {\cal B}_n$, we start with a
monotone network $\beta$ which computes $f$. We have no
knowledge about the structure of $\beta$. In particular, we have no knowledge about the DNF-representations 
of the functions computed at the nodes of $\beta$. Let $g_1,g_2, \ldots,g_t$ be the nodes of $\beta$ numbered
in any topological order. Starting with $g_1$, the DNF-representations
$\mbox{DNF}_{\beta}(g_i)$, $1 \leq i \leq t$ are treated in this order. The idea is to replace
$\mbox{DNF}_{\beta}(g_i)$ by an approximation $\mbox{DNF'}_{\beta}(g_i)$ such that we have the needed structural
information. After the replacement, $\mbox{DNF}_{\beta}(g_j)$, $j > i$ has to be updated such that for its
construction $\mbox{DNF'}_{\beta}(g_i)$ is used instead of $\mbox{DNF}_{\beta}(g_i)$. Therefore, not the
function $f$ but an approximation $f'$ of $f$ is computed at the output node of $\beta$. Hence, there are inputs
$c \in \{0,1\}^n$ such that $f'(c) \not= f(c)$. Let $g_i$ be the last node for which $\mbox{DNF}_{\beta}(g_i)$
has been replaced by $\mbox{DNF'}_{\beta}(g_i)$. In the subsequence, $\mbox{DNF}_{\beta}(g_j)$ denotes for $j > i$
the DNF-representation of the current function computed at the node $g_j$ and for $j \leq i$,
$\mbox{DNF}_{\beta}(g_j)$ immediately before its replacement.

Let $f_1$ ($f_2$) denote the function computed at the output node $g_t$ after the approximation of
$\mbox{DNF}_{\beta}(g_{i-1})$ ($\mbox{DNF}_{\beta}(g_i)$) and before the approximation of
$\mbox{DNF}_{\beta}(g_i)$ ($\mbox{DNF}_{\beta}(g_{i+1})$). We say that the approximator of the node $g_i$
{\em introduces an error\/} with respect to the input $c \in \{0,1\}^n$ if $f_1(c) = f(c)$ but
$f_2(c) \not= f(c)$. Note that $f'(c) \not= f(c)$ implies that there exists a node $g_i$ in $\beta$ such that
the approximator of $g_i$ introduces an error with respect to the input $c$. The approximators should be designed
in a way such that the following is fulfilled:
\begin{enumerate}
\item
  After the replacement of $\mbox{DNF}_{\beta}(g_t)$, the number of inputs $c \in \{0,1\}^n$ with
  $f'(c) \not= f(c)$ is ``large''.
\item
  For all nodes $g_i$, $1 \leq i \leq t$, the number of inputs $c \in \{0,1\}^n$ where an error with respect to
  $c$ is introduced by the replacement of $\mbox{DNF}_{\beta}(g_i)$ by $\mbox{DNF'}_{\beta}(g_i)$ is ``small''.
\end{enumerate}
Note that these properties imply that a monotone network computing the function $f$ has to contain ``many'' gates.
How to approximate DNF-formulas such that these properties are fulfilled?

The general idea is to bound the size of the monomials in the DNF-formulas constructed at the nodes in $\beta$.
The size of a monomial can be its length; i.e., its number of distinct literals, or another measure.
Let $r$ be the upper bound for the size of a monomial in an approximator with respect to a node $g_i$. An obvious
way to bound the size of the monomials would be the following:
\begin{itemize}
\item
  For the construction of the approximator $\mbox{DNF'}_{\beta}(g_i)$ construct the DNF-representation
  $\mbox{DNF}_{\beta}(g_i)$ of the current function computed at $g_i$ and remove each monomial of size larger
  than $r$.
\end{itemize}
The effect of the removal of monomials from $\mbox{DNF}_{\beta}(g_i)$ to the DNF-represen\-ta\-tion
$\mbox{DNF}_{\beta}(g_t)$ of $\mbox{res}_{\beta}(g_t)$ is the removal of some monomials in $\mbox{DND}_{\beta}(g_t)$.
Hence, an error could be introduced only for inputs $c \in f^{-1}(1)$. Next we describe the construction of the
approximators more in detail.

For input nodes, the approximator and the original DNF-representation are the same. For the comparision of
the approximator $\mbox{DNF'}_{\beta}(g_i)$ and the DNF-representation $\mbox{DNF}_{\beta}(g_i)$ of the current
function computed at $g_i$ immediately before the approximation of $\mbox{DNF}_{\beta}(g_i)$ suppose that $g_{i_1}$
and $g_{i_2}$ are the direct predecessors of the gate $g_i$. By construction, each monomial in
$\mbox{DNF'}_{\beta}(g_{i_1}) = \bigvee_{j=1}^{t_1} m_j$ and in $\mbox{DNF'}_{\beta}(g_{i_2}) = \bigvee_{l=1}^{t_2} m'_l$
has size at most $r$. Therefore, if $g_i$ is an $\vee$-gate, each monomial in the DNF-representation
$\mbox{DNF}_{\beta}(g_i)$ of the current function computed at $g_i$ has size at most $r$. Hence, we define
$\mbox{DNF'}_{\beta}(g_i) := \mbox{DNF}_{\beta}(g_i)$. No error is introduced by the approximator
$\mbox{DNF'}_{\beta}(g_i)$. But the number of monomials in $\mbox{DNF'}_{\beta}(g_i)$ could be the double   
of the number of monomials in $\mbox{DNF'}_{\beta}(g_{i_1})$ or in $\mbox{DNF'}_{\beta}(g_{i_2})$. 
If $g_i$ is an $\wedge$-gate then
$$
\mbox{DNF}_{\beta}(g_i) = \bigvee_{j=1}^{t_1}\bigvee_{l=1}^{t_2} (m_j \wedge m'_l).
$$
To obtain $\mbox{DNF'}_{\beta}(g_i)$, we remove from $\mbox{DNF}_{\beta}(g_i)$ all monomials $m_jm'_l$ of size
larger than $r$. To get a large lower bound for the function $f \in {\cal B}_n$, the function $f$ must have the
following property:
\begin{itemize}
\item[{\bf F1}]
  Only ``few'' inputs in $f^{-1}(1)$ fulfill a monomial of size larger than $r$.
\end{itemize}

If the number of monomials which are removed would be small enough then perhaps, we could prove an upper bound for
the number of errors introduced by the approximation at an $\wedge$-gate which is small enough.
We need a mechanism which bounds the number of monomials removed during the construction of an approximator. Two
such mechanisms are known,
{\em CNF-DNF-approximators} which switch between CNF- and DNF-formulas and approximators which use the sunflower
lemma discovered by Erd\H{o}s and Rado \cite{ErRa}. We call such an approximator {\em sunflower-approximator}.
Next, we review both approximators with respect to their use in monotone networks.

\section{Approximators in Monotone Networks}

Let $f \in {\cal B}_n$ be the monotone function for which we intend to prove a large lower bound of its
monotone complexity. Let $\beta = g_1,g_2, \ldots,g_t$ be a monotone network which computes $f$ at its output
node $g_t$. To get a large lower bound, the approximation method has to take care that the following
property is fulfilled:
\begin{itemize}
\item[{\bf A1}]
  Only ``few'' monomials are removed to obtain $\mbox{DNF'}_{\beta}(g_i)$ from $\mbox{DNF}_{\beta}(g_i)$.
\end{itemize}
The methods to obtain this property in CNF-DNF- and in sunflower-approxi\-ma\-tors are different. 
Since CNF-DNF-approximators are less difficult, we describe these approximators first.

\subsection{CNF-DNF-Approximators}

CNF-DNF-approximators are introduced implicitly by Haken \cite{Ha} and explicitly by Jukna \cite{Ju}, Berg and
Ulfberg \cite{BeUl} and Amano and Maruoka \cite{AmMa}.
To understand the idea of CNF-DNF-approximators let us consider the organization of a CNF/DNF-switch as the
construction of a tree $T$ as described in Section 2. Obviously, the outdegree of an inner node $w$ cannot be
larger than the number of literals in the clause which is expanded at the leaf $w$ during the construction of the
tree $T$. Let $m(w)$ denote the monomial associated with the path from the root to the node $w$. After the
performance of the CNF/DNF-switch consider any path $P$ from the root to a leaf $w$ in $T$. Let $v$ be any node
on $P$. Obviously, the monomial $m(v)$ is a prefix of the monomial of $m(w)$.
If we can ensure that on each edge starting from a node with outdegree at least two, the size of the
corresponding monomial increases by one then the number of different prefixes of size exactly $r$ 
could be bounded by $l(k)^r$ where $l(k)$ is the maximal number of literals in a clause expanded during the
CNF/DNF-switch. $k$ will be an upper bound for the size of a clause used in the construction of an approximator.
After the construction of $T$, all monomials of size larger than $r$ are removed. Each such a monomial has a prefix
of size $r$ which has to be fulfilled by each input which fulfills the monomial. Therefore, an upper bound for the
number of different such prefixes can be used in a lower bound proof. We call a CNF/DNF-switch followed by the
elimination of all monomials of size larger than $r$ an {\em CNF/DNF-approximator switch\/}. Analogously, if $l(r)$
is the maximal number of literals in a monomial of size at most $r$  expanded during a DNF/CNF-switch then we
obtain an upper bound of $l(r)^k$ for the number of different prefixes of size exactly $k$. After the construction
of $T$, all clauses of size larger than $k$ are removed. Such a switch is called {\em DNF/CNF-approximator switch}.  

This observation yields the idea to approximate with respect to each node $g_i$ in $\beta$
$\mbox{DNF}_{\beta}(g_i)$ by a DNF-formula $\mbox{DNF'}_{\beta}(g_i)$ which contains only monomials of size at most
$r$ and also $\mbox{CNF}_{\beta}(g_i)$ by a CNF-formula $\mbox{CNF'}_{\beta}(g_i)$ which contains only clauses of
size at most $k$. Note that by the removal of clauses from $\mbox{CNF}_{\beta}(g_i)$ only for inputs
$c \in f^{-1}(0)$ an error could be introduced.
To get a large lower bound for the monotone complexity of $f$ using CNF-DNF-approximators, the function $f$ must
have the following additional property:
\begin{itemize}
\item[{\bf F2}]
  Only ``few'' inputs in $f^{-1}(0)$ falsify a clause of size larger than $k$.
\end{itemize}

To get a large lower bound for the monotone complexity of $f$, the number of inputs for which
$\mbox{CNF'}_{\beta}(g_t)$ or $\mbox{DNF'}_{\beta}(g_t)$ compute the wrong value has to be ``large''. Therefore,
the function $f$ must have the following additional property:
\begin{itemize}
\item[{\bf F3}]
  At least one of the following two properties is fulfilled:
  \begin{enumerate}
\item
  If $\mbox{CNF'}_{\beta}(g_t)$ is not the constant function one then ``many'' inputs in $f^{-1}(1)$ do not
  fulfill $\mbox{CNF'}_{\beta}(g_t)$.
\item
  If $\mbox{DNF'}_{\beta}(g_t)$ is not the constant function zero then ``many'' inputs in $f^{-1}(0)$ do not
  falsify $\mbox{DNF'}_{\beta}(g_t)$.
  \end{enumerate}
\end{itemize}

Instead of using the whole sets $f^{-1}(1)$ and $f^{-1}(0)$, more appropriate subsets $T_1 \subseteq f^{-1}(1)$
and $T_0 \subseteq f^{-1}(0)$ could be used to prove the lower bound.
Now we are prepared to give a precise description of CNF-DNF-approxi\-ma\-tors. We distinguish three cases.

\medskip
\noindent
{\em Case 1:\/} $g_i$ is an input node.

\medskip
Then
$$
\mbox{CNF'}_{\beta}(g_i) := \mbox{CNF}_{\beta}(g_i) \; \mbox{ and } \;
\mbox{DNF'}_{\beta}(g_i) := \mbox{DNF}_{\beta}(g_i).
$$
Obviously, no error is introduced by both approximators.

\medskip
\noindent
 {\em Case 2:\/} $g_i$ is an $\wedge$-gate with direct predecessors $g_{i_1}$ and $g_{i_2}$.

\medskip
Then
$$
\mbox{CNF'}_{\beta}(g_i) := \mbox{CNF'}_{\beta}(g_{i_1}) \wedge \mbox{CNF'}_{\beta}(g_{i_2}).
$$
Since the size of each clause in $\mbox{CNF'}_{\beta}(g_{i_1})$ and in $\mbox{CNF'}_{\beta}(g_{i_2})$ is at most $k$,
each clause in $\mbox{CNF'}_{\beta}(g_i)$ has also at most size $k$. Since
$\mbox{CNF'}_{\beta}(g_i) = \mbox{CNF}_{\beta}(g_i)$, no error is introduced by the approximation.

$\mbox{DNF'}_{\beta}(g_i)$ is obtained from $\mbox{CNF'}_{\beta}(g_i)$ by performing a CNF/DNF-approx\-i\-ma\-tor
switch.

\medskip
\noindent
{\em Case 3:\/} $g_i$ is an $\vee$-gate with direct predecessors $g_{i_1}$ and $g_{i_2}$.

\medskip
Then
$$
\mbox{DNF'}_{\beta}(g_i) := \mbox{DNF'}_{\beta}(g_{i_1}) \vee \mbox{DNF'}_{\beta}(g_{i_2}).
$$
Since the size of each monomial in $\mbox{DNF'}_{\beta}(g_{i_1})$ and in $\mbox{DNF'}_{\beta}(g_{i_2})$ is at most $r$,
each monomial in $\mbox{DNF'}_{\beta}(g_i)$ has also at most size $r$. Since
$\mbox{DNF'}_{\beta}(g_i) = \mbox{DNF}_{\beta}(g_i)$, no error is introduced by the approximation.

$\mbox{CNF'}_{\beta}(g_i)$ is obtained from $\mbox{DNF'}_{\beta}(g_i)$ by performing a DNF/CNF-approx\-i\-mator
switch.

\medskip
For the application of CNF-DNF-approximators to a monotone function $f \in {\cal B}_n$, to obtain a large lower
bound
for its monotone complexity, it has to exists large sets $T_1 \subseteq f^{-1}(1)$ and $T_0 \subseteq f^{-1}(0)$ of
inputs such that the following properties are fulfilled:
\begin{enumerate}
\item
  There is a ``small'' upper bound $n_1$ for the number of inputs in $T_1$ which fulfill any monomial of size
  $r+1$.  
\item
  There is a ``small'' upper bound $n_0$ for the number of inputs in $T_0$ which falsify any clause of size $k+1$.  
\item
  At least one of the following two cases is fulfilled:
  \begin{enumerate}
  \item
    There is a constant $d_1 < 1$ such that a clause of size $k$ is fulfilled by at most $d_1|T_1|$ inputs in $T_1$.
  \item
    There is a constant $d_0 < 1$ such that a monomial of size $r$ is falsified by at most $d_0|T_0|$ inputs in $T_0$.
  \end{enumerate}
\end{enumerate}

\medskip
To get a lower bound for the monotone complexity of $f$, the properties are used in the following way. Assume that
the constant $d_1$ exists. With respect to $\mbox{CNF'}_{\beta}(g_t)$, two situations can arise. If
$\mbox{CNF'}_{\beta}(g_t)$ computes the constant function one then for no input $c \in T_0$, the value $f(c)$ is
computed correctly. Therefore, for each input $c \in T_0$ there is an $\vee$- gate $g_i$ such that the construction
of $\mbox{CNF'}_{\beta}(g_i)$ introduce an error with respect to the input $c$. Since at an $\vee$-gate, an error for
at most $l(r)^k n_0$ inputs in $T_0$ is introduced, we obtain the lower bound $\frac{|T_0|}{l(r)^kn_0}$ for the
monotone complexity of $f$. Otherwise, $\mbox{CNF'}_{\beta}(g_t)$ contains a non-empty clause of size at most $k$.
Therefore by the third property, $\mbox{CNF'}_{\beta}(g_t)$ can be fulfilled by at most $d_1|T_1|$ inputs in $T_1$.
Since at an $\wedge$-gate, an error for at most $l(k)^r n_1$ inputs in $T_1$ is introduced, we obtain the lower
bound $\frac{(1-d_1)|T_1|}{l(k)^rn_1}$ for the monotone complexity of $f$. The case that $d_0$ exists can be
discussed analogously. 

\subsection{Sunflower-Approximators}

To bound the number of monomials which could be removed during the construction of an approximator,
sunflower-approximators use the sunflower lemma or a modification of the sunflower lemma. Since
CNF-DNF-approxi\-ma\-tors do not use such a combinatorial
lemma, they seem to be simpler than sunflower-approximators. But the properties which a Boolean function must have
to get a large lower bound for its monotone complexity using the methods are different. Both methods use the
following property:
\begin{itemize}
\item[{\bf F1}]
  Only ``few'' inputs in $f^{-1}(1)$ fulfill a monomial of size larger than $r$.
\end{itemize}
The properties F2 and F3 needed if we use CNF-DNF-approximators are replaced by two other properties. Therefore,
the sets of functions for which the application of the methods results into a large lower bound for the monotone
complexity may be different. Note that with respect to the perfect matching function, we know a proof of a
super-polynomial lower bound which uses a sunflower-approximator \cite{Ra3} but no such a proof which uses a
CNF-DNF-approximator.
First of all, we will review the sunflower lemma and its use for proving a lower bound.

\medskip
The {\em sunflower lemma} of Erd\H{o}s and Rado \cite{ErRa} is the central combinatorial property to bound the
number of monomials removed during the construction of an approximator. The basis of our description is the
excellent presentation of Jukna in \cite{Ju2,Ju3}.

A {\em sunflower\/} with $p$ {\em pedals\/} and {\em core\/} $T$ is a collection $S_1, S_2, \ldots,S_p$ of $p$ sets
such that $S_i \cap S_j = T$ for $1 \leq i < j \leq p$. Note that $p$ pairwise disjoint sets is a sunflower with
empty core. The following sunflower lemma means that each family of nonempty sets which is large enough must
contain a sunflower with $p$ pedals.

\begin{lem}
  Let ${\cal F}$ be a family of non-empty sets each of size at most $r$. If $|{\cal F}| > r!(p-1)^r$ then ${\cal F}$
  contains a sunflower with $p$ pedals.
\end{lem}

If we relax the property that the core $T$ lies entirely in all sets $S_1, \ldots,S_p$ such that the
differences $S_i \setminus T$, $1 \leq i \leq p$ are non-empty and mutually disjoint then we obtain a lemma proved
by F\"uredi in 1978 \cite{Fu}. The {\em common part} of $p$ distinct finite sets $S_1, S_2, \ldots,S_p$ is the set
$T := \bigcup_{i \not= j} (S_i \cap S_j)$. 

\begin{lem}
  Let ${\cal F}$ be a family of non-empty sets each of size at most $r$. If $|{\cal F}| > (p-1)^r$ then ${\cal F}$
  contains $p$ sets with common part of size less than $r$.
\end{lem}

Razborov's approximator \cite{Ra1,Ra3} are based on the sunflower lemma and on F\"uredi's lemma. Andreev
\cite{An,An2} uses his own modification of the sunflower lemma. No matter which modification of the sunflower
lemma is used, the essential properties of the approximators are the same. Therefore, we only describe
approximators which uses the sunflower lemma directly.

To use the sunflower lemma, a set $S(m)$ of the same size as $m$ is constructed for each monomial $m$. For example,
if its length is the size of $m$, $S(m)$ is the set of all variables in $m$. The set $S(m)$ corresponds to the
monomial $m$. The idea is to use an appropriate $r$
as the upper bound for the size of a monomial and $l := r!(p-1)^r$ as the upper bound for the number of monomials
in the approximator $\mbox{DNF'}_{\beta}(g_i)$, $1 \leq i \leq t$. For the construction of the approximators, the
nodes in $\beta$ are considered in a topological order such that the approximators of the direct predecessors
$g_{i_1}$ and $g_{i_2}$ are already constructed when $\mbox{DNF'}_{\beta}(g_i)$ is constructed. Note
that all monomials in $\mbox{DNF'}_{\beta}(g_{i_1})$ and in $\mbox{DNF'}_{\beta}(g_{i_2})$ have size at most $r$ and the
number of monomials in both approximators is at most $l$.

After the construction of $\mbox{DNF}_{\beta}(g_i)$ where $g_i$ is an $\vee$-gate, the number of monomials in
$\mbox{DNF}_{\beta}(g_i)$ can exceed the upper bound $l$ but is at most $2l$. Then, by the sunflower lemma, there
are $p$ monomials $m_1, m_2, \ldots, m_p$ such that the corresponding sets $S(m_1), S(m_2), \ldots, S(m_p)$ form a
sunflower. The core $T$ of the sunflower corresponds to a monomial $m(T)$. Then in $\mbox{DNF}_{\beta}(g_i)$, the
monomials $m_1, m_2, \ldots, m_p$ are replaced by the single monomial $m(T)$ which is a submonomial of each monomial
$m_j$, $1 \leq j \leq p$. This operation is called a {\em plucking\/}. The effect of a plucking to
$\mbox{DNF}_{\beta}(g_t)$ is the replacement of some monomials by a proper submonomial. Hence, a plucking can only
introduce an error for inputs in $f^{-1}(0)$. As long as possible pluckings are performed leading to the approximator
$\mbox{DNF'}_{\beta}(g_i)$. By the sunflower lemma, the number of monomials in $\mbox{DNF'}_{\beta}(g_i)$ is at most $l$.
Since at the beginning, the number of monomials is at most $2l$, less than $2l$ pluckings are performed.

After the construction of $\mbox{DNF}_{\beta}(g_i)$ where $g_i$ is an $\wedge$-gate, the size of some monomials can
exceed the upper bound $r$. We remove all these monomials first. Since $\mbox{DNF'}_{\beta}(g_{i_1})$ and also
$\mbox{DNF'}_{\beta}(g_{i_1})$ contain at most $l$ monomials, at most $l^2$ monomials are removed. Then, the plucking
procedure is applied to the
remaining monomials obtaining the approximator $\mbox{DNF'}_{\beta}(g_i)$. Since the approximators of the direct
predecessors of $g_i$ contain at most $l$ monomials, at most $l^2$ pluckings are performed.

For the application of sunflower-approximators to a monotone function $f \in {\cal B}_n$, to obtain a large lower
bound for its monotone complexity, it has to exist large sets $T_1 \subseteq f^{-1}(1)$ and $T_0 \subseteq f^{-1}(0)$
of inputs such that the following properties are fulfilled:
\begin{enumerate}
\item
  There is a ``small'' upper bound $n_1$ for the number of inputs in $T_1$ which fulfill any monomial of size larger
  than $r$.
\item
  There is a ``small'' upper bound $n_0$ for the number of inputs in $T_0$ for which an error is introduced because
  the performance of a plucking.
\item
  At least one of the following two cases is fulfilled:
  \begin{enumerate}
  \item
    There is a constant $d_1 < 1$ such that $\mbox{DNF'}_{\beta}(g_t)$ computes the constant function one or
    $\mbox{DNF'}_{\beta}(g_t)$ is satisfied by at most $d_1|T_1|$ inputs in $T_1$.
  \item
    There is a constant $d_0 < 1$ such that $\mbox{DNF'}_{\beta}(g_t)$ computes the constant function zero or
    $\mbox{DNF'}_{\beta}(g_t)$ is falsified by at most $d_0|T_0|$ inputs in $T_0$.
  \end{enumerate}
\end{enumerate}

To get a lower bound for the monotone complexity of the function $f$, sunflower approximators are used in the
following way. Assume that the constant $d_1$ exists. If $\mbox{DNF'}_{\beta}(g_t)$ computes the constant function
one then for no input $c \in T_0$, the value $f(c)$ is computed correctly. Only pluckings introduce an error for
an input in $T_0$. Since a single plucking introduces an error for at most $n_0$ inputs in $T_0$, at least
$\frac{|T_0|}{n_0}$ pluckings are performed. Since at most $l^2$ pluckings are performed at a gate in $\beta$, we
obtain the lower bound $\frac{|T_0|}{l^2n_0}$ for the monotone complexity of $f$. Otherwise, there are $(1-d_1)|T_1|$
inputs in $T_1$ which do not satisfy $\mbox{DNF'}_{\beta}(g_t)$. Only the removal of a monomial at an $\wedge$-gate
can introduce an error for an input in $T_1$. The removal of one monomial of size larger than $r$ can introduce an
error for at most $n_1$ inputs in $T_1$. Hence, at least $\frac{(1-d_1)|T_1|}{n_1}$ monomials are removed. Since at
most $l^2$ monomials are removed at an $\wedge$-gate, we obtain the lower bound $\frac{(1-d_1)|T_1|}{l^2n_1}$ for
the monotone complexity of $f$. The case that $d_0$ exists can be discussed analogously.

\section{The Extension of the Approximation Method} 

Our goal is to extend the approximation method such that it can be used to prove a super-polynomial lower bound for
the standard complexity of an appropriate Boolean function $f \in {\cal B}_n$ or to understand why this is not
possible. Let $\beta = g_1, g_2, \ldots, g_t$ be a standard network which computes a function $f \in {\cal B}_n$ at
its output node $g_t$. As in monotone networks our goal is to approximate the DNF-formulas constructed at the nodes
in $\beta$; i.e., we replace
$\mbox{DNF}_{\beta}(g_i)$ by an approximator $\mbox{DNF'}_{\beta}(g_i)$. Exactly as in monotone networks, we define
the notion that $\mbox{DNF'}_{\beta}(g_i)$ {\em intoduces an error with respect to the input $c \in \{0,1\}^n$}.
Again, the general idea is to bound the size of the monomials in the DNF-formulas constructed at the nodes of
$\beta$. In contrast to monotone networks, a monomial in $\mbox{DNF}_{\beta}(g_i)$ can contain both kinds of
literals. Hence, with respect to the definition of the size of a monomial, we have two possibilities:
\begin{enumerate}
\item
  The size of a monomial depends on both kinds of literals.
\item
  The size of a monomial depends only on one kind of literals.
\end{enumerate}
Before discussing both cases, let us review the needed properties of the function $f$ such that a large lower
bound could be proved for $f$ using the approximation method. Since the approximation method obtains
$\mbox{DNF'}_{\beta}(g_i)$ by the removal of all monomials of size larger than a given bound $r$, the first property
is that the number of inputs $c \in T_1$ which fulfill a monomial of size larger than $r$ is small enough. Since
at the output node $g_t$, the value of many inputs $c \in T_1 \cup T_0$ has to be computed incorrectly, the second
property is that the number of inputs $c \in T_1 \cup T_0$ with $\mbox{DNF'}_{\beta}(g_t)(c) \not= f(c)$ is large
enough.
The approximation method takes care that the number of monomials which are removed from $\mbox{DNF}_{\beta}(g_i)$ to
obtain $\mbox{DNF'}_{\beta}(g_i)$ is small enough. This is done in the following way:

\smallskip
\noindent
{\em CNF-DNF-approximators:}

\smallskip
At each gate $g_i$, the CNF-formula $\mbox{CNF}_{\beta}(g_i)$ is approximated by a CNF-formula $\mbox{CNF'}_{\beta}(g_i)$
as well. $\mbox{CNF'}_{\beta}(g_i)$ is obtained from $\mbox{CNF}_{\beta}(g_i)$ by the removal of all clauses of larger
size than a given bound $k$. Therefore, the additional property that the number of inputs $c \in T_0$ which falsify a
clause of size larger than $k$ is small enough is needed.
In dependence of $r$ and $k$, upper bounds $h_m(r)$ and $h_c(k)$ of the number of variables
in a monomial of size at most $r$ and in a clause of size at most $k$, respectively are derived. Then, upper bounds
$h_1(k)^r$ and $h_2(r)^k$ for the number of different prefixes of size exactly $r$ of the monomials removed at an
$\wedge$-gate $g_i$ and the number of different prefixes of size exactly $k$ of the clauses removed
at an $\vee$-gate $g_i$ are estimated. To get a large lower bound, $p := h_1(k)$ has to be non-constant.

\smallskip
\noindent
{\em Sunflower-approximators:}

\smallskip
To bound the number of monomials removed from $\mbox{DNF}_{\beta}(g_i)$ to obtain $\mbox{DNF'}_{\beta}(g_i)$,
sunflower-approximators use parameters $2 \leq r,p \leq n$ where $r$ is an given upper bound for the size of the
monomials in $\mbox{DNF'}_{\beta}(g_i)$ and $p$ is the number of pedals with respect to the sunflower lemma.
Using the sunflower lemma, the number of monomials in the approximators is bounded by $l := r!(p-1)^r$. Hence,
an upper bound of $l^2$ for the number of monomials removed at an $\wedge$-gate $g_i$ is obtained.
To get a large lower bound, $p$ has to be non-constant.

\smallskip
Now we will discuss the case that the size of a monomial depends on both kind of literals. We will give some evidence
that in this case, the approximation method cannot be extended to prove a super-linear lower bound for the standard
complexity of any Boolean function $f \in {\cal B}_n$. 

As described above, with respect to each known approximation method, the upper bound for the number of monomials removed
at an $\wedge$-gate $g_i$ for obtaining $\mbox{DNF'}_{\beta}(g_i)$ from $\mbox{DNF}_{\beta}(g_i)$ is at least $p^r$ where $r$
is the upper bound for the size of the monomials in $\mbox{DNF'}_{\beta}(g_i)$ and $p$ is non-constant. In the case that
both kinds of literals are approximated, such an upper bound seems to be too large.
We will explain this for sizes which depend on the number of negated and the number of non-negated variables in a monomial
$m$. Let $r_0$ ($r_1$) be the upper bound for the number of negated (non-negated) variables of a monomial $m$ in
$\mbox{DNF'}_{\beta}(g_i)$; i.e., each monomial which does not fulfill both bounds is removed from $\mbox{DNF}_{\beta}(g_i)$.
Let $r := r_0 + r_1$. Note that each monomial of length larger than $r$ cannot fulfill both bounds. The following lemma
shows that that $2^r$ monomials of length $r$ are sufficient such that each input $c \in f^{-1}(1)$ fulfills at least one
of these monomials.

\begin{lem}
  Let $f$ be any Boolean function in ${\cal B}_n$. There are $2^r$ monomials of length $r$ such that each input
  $c \in f^{-1}(1)$ fulfills at least one of these monomials.
\end{lem}
{\bf Proof}:
Fix any $r$ variables $x_{i_1},x_{i_2}, \ldots,x_{i_r}$. Consider any $c \in f^{-1}(1)$. Let
$$
m'_c := y_{i_1}y_{i_2} \ldots y_{i_r}
$$
where for $1 \leq j \leq r$
$$
y_{i_j} := \left\{ \begin{array}{ll}
    x_{i_j} & \mbox{if $c_{i_1} = 1$} \\
    \neg x_{i_j} & \mbox{if $c_{i_j} =0$}
  \end{array}
  \right.
$$
By construction, $m'_c(c) = 1$ and $|m'_c| = r$. There are $2^r$ monomials which use exactly the variables
$x_{i_1},x_{i_2}, \ldots,x_{i_r}$.  
$\Box$

\smallskip
Note that each submonomial of $m'_c$ is fulfilled by $c$ as well. 
At most $2^r$ monomials of length at most $r$ could suffice such that their removal could introduce an error for
each input $c \in f^{-1}(1)$. Since $p^r > 2^r$ for non-constant $p$, the approximation of the DNF-formula with respect to
one $\wedge$-gate could destroy the correct computation of $f(c)$ for all $c \in f^{-1}(1)$.
This shows that the first property seems not be fulfilled if both kind of literals are approximated.

It remains the consideration of the case that the size of a monomial depends only on one kind of literals.
We discuss CNF-DNF-approximators first.

\subsection{Extended CNF-DNF-Approximators}

We consider the subcase that the non-negated variables are approximated. The other subcase can be
discussed in the same way.
Let $\beta = g_1,g_2, \ldots, g_t$ be a standard network which computes a Boolean function $f \in {\cal B}_n$
at its output node $g_t$.
The sizes of a monomial $m$ or of a clause $d$ depend only on its non-negated variables. This implies
that any number of negated variables can be contained in each monomial $m$ in $\mbox{DNF'}_{\beta}(g_i)$ and
also in each clause $d$ in $\mbox{CNF'}_{\beta}(g_i)$. Since $\beta$ computes the function $f$ at its output
node $g_t$, before any approximation, $\mbox{DNF}_{\beta}(g_t)$ contains for each $c \in f^{-1}(1)$ a monomial
$m_c$ such that $m_c(c) = 1$. Only for $c = (1,1, \ldots,1)$, we can exclude that $m_c$ contains any negated
variable. Furthermore, $\mbox{CNF}_{\beta}(g_t)$ contains for each $c \in f^{-1}(0)$ an $f$-clause $d_c$ such that
$d_c(c) = 0$. Before any approximation, each $f$-clause $d$ in $\mbox{CNF}_{\beta}(g_t)$ contains a literal in
$m_c$ for each $c \in f^{-1}(1)$.

We have to define large sets $T_1 \subseteq f^{-1}(1)$ and
$T_0 \subseteq f^{-1}(0)$ of inputs such that the following properties are fulfilled:
\begin{enumerate}
\item
  There is a ``small'' upper bound $n_1$ for the number of inputs in $T_1$ which fulfill any monomial of size
  $r+1$.  
\item
  There is a ``small'' upper bound $n_0$ for the number of inputs in $T_0$ which falsify any clause of size $k+1$.  
\item
  At least one of the following two cases is fulfilled:
  \begin{enumerate}
  \item
    There is a constant $d_1 < 1$ such that a clause of size $k$ is fulfilled by at most $d_1|T_1|$ inputs in $T_1$.
  \item
    There is a constant $d_0 < 1$ such that a monomial of size $r$ is falsified by at most $d_0|T_0|$ inputs in $T_0$.
  \end{enumerate}
\end{enumerate}
But without any further information about the structure of the network it seems to be impossible to fulfill the
third property without destroying the first or the second property. Since the third property divides into two
cases, we have to consider the two situations that $\mbox{CNF'}_{\beta}(g_t)$ is not the constant function one and
that $\mbox{DNF'}_{\beta}(g_t)$ is not the constant function zero.

\medskip
If $\mbox{CNF'}_{\beta}(g_t)$ is not the constant function one then $\mbox{CNF'}_{\beta}(g_t)$ must contain at
least one clause $d$. Note that $d$ is a subclause of any $f$-clause $d'$. Let $\mbox{pc}(d)$ be a prime clause
which is a subclause of the $f$-clause $d'$. 
For each $c \in  f^{-1}(1)$ such that $c$ fulfills a literal contained in $d$, it cannot be excluded that $f(c)$
is computed correctly by $\mbox{CNF'}_{\beta}(g_t)$. Therefore, for each $c \in f^{-1}(1)$ such that $d$ is not
satisfied by $c$, $c_j = 1$ if $\neg x_j$ is contained in the clause $d$. $d$ can contain each $\neg x_j$ where
the variable $x_j$ is not contained in $\mbox{pc}(d)$. To get the property that a clause of size $k$ could be
fulfilled by at most $d_1|T_1|$ inputs in $T_1$, the set $T_1$ should only contain inputs $c \in f^{-1}(1)$ such
that there is a prime clause $d(c)$ with $c_j = 1$ for all $x_j$ not contained in $d(c)$. But for such an input
$c \in T_1$, a monomial of size $r+1$ which is fulfilled by $c$ must not be a
submonomial of a prime implicant which is fullfilled by $c$. Therefore, a small upper bound $n_1$ for the
number of inputs in $T_1$ which fulfill any monomial of size $r+1$ cannot be proved in the usual way.
Note that usually, the set $T_1$ contains exactly those inputs which correspond to the prime implicants of the
function. Because of the structure of the prime clauses of the considered functions, I have found no other way
to prove a small upper bound.

If $\mbox{DNF'}_{\beta}(g_t)$ is not the constant function zero then $\mbox{DNF'}_{\beta}(g_t)$ must contain at
least one monomial $m$. Note that $m$ is a submonomial of an implicant $m'$ of $f$. Let $\mbox{pi}(m)$ be a
prime implicant which is a submonomial of the implicant $m'$.
For each $c \in  f^{-1}(0)$ such that $c$ falsifies a literal contained in $m$, it cannot be excluded that $f(c)$
is computed correctly by $\mbox{DNF'}_{\beta}(g_t)$. Therefore, for each $c \in f^{-1}(0)$ such that $m$ is
satisfied by $c$, $c_j = 0$ if $\neg x_j$ is contained in the monomial $m$.
$m$ can contain each $\neg x_j$ where the variable $x_j$ is not
contained in $\mbox{pi}(m)$. To get the property that a monomial of size $r$ could be
falsified by at most $d_0|T_0|$ inputs in $T_0$, the set $T_0$ should only contain inputs $c \in f^{-1}(0)$ with
the property that there is a prime implicant $p(c)$ such that $c_j = 0$ for all $x_j$ not contained
in $p(c)$. For such a set $T_0$, a small upper bound $n_0$ for the number of inputs in $T_0$ which falsify any
clause of size $k+1$ cannot be proved in the usual way. Because of the structure of the prime implicants of the
considered functions, I have found no other way to prove a small upper bound.

This gives evidence that with respect to CNF-DNF-approximators, the needed properties cannot be fulfilled if only
one kind of literals is approximated and no other properties of the network are used.

Altogether, we have given some evidence that extended CNF-DNF-ap\-prox\-i\-mators alone cannot be used to prove a
lower bound for the standard complexity of any Boolean function $f \in {\cal B}_n$.

\subsection{Extended Sunflower-Approximators}

Now, we shall investigate the expandability of sunflower-approximators. Unlike CNF-DNF-approxi\-ma\-tors which
have an obvious extension, we have to explain the extension of sunflower-approximators.
Given any standard network $\beta$ computing the considered function $f \in {\cal B}_n$, we will use extended
sunflower-approximators for the approximation of $\mbox{DNF}_{\beta}(g)$ where $g$ is a node in $\beta$. Hence, we
separate in each monomial $m_j$ of $\mbox{DNF}_{\beta}(g)$ the negated and the non-negated variables obtaining
$$
m_j = m_{j_0}m_{j_1}
$$
where $m_{j_0}$ contains exactly the negated variables and $m_{j_1}$ contains exactly the non-negated variables
of $m_j$. If $m_j$ does not contain any negated (non-negated) variable then $m_{j_0} = \varepsilon$
($m_{j_1} = \varepsilon$). After doing this, we write for $\mbox{DNF}_{\beta}(g) = \bigvee_{j=1}^s m_j$ 
$$
\mbox{DNF}_{\beta}(g) = \bigvee_{j=1}^s m_{j_0}m_{j_1}.
$$

We describe the extension of sunflower-approximators for the case that the non-negated variables are approximated. The
approximation of the negated variables can be done analogously. The construction is more complicated than in the
monotone case.  The difficulties come from the incorporation of the negated part of the monomials which we do not
approximate. 
We consider the case that the {\em size\/} of the non-negated part $m_1$ of a monomial $m = m_0m_1$ is its length;
i.e., the number of different variables in $m_1$. For the application of the sunflower lemma, we use for each monomial
$m = m_0m_1$ the set $V(m_1)$ of variables contained in $m_1$. 

Let $g_1,g_2, \ldots, g_t$ be the nodes of $\beta$ numbered in any topological order. In dependence of the
approximators corresponding to the direct predecessors of the considered node $g_i$ and the operation $op_i$ of $g_i$,
we define the approximator of $\mbox{DNF}_{\beta}(g_i)$. ${\cal M}_n$ ($\overline{{\cal M}}_n$) denotes the set of
monomials consisting of only non-negated (negated variables) from $V_n$ including the empty monomial. $\overline{m}$
denotes always a monomial in $\overline{{\cal M}}_n$.
First of all, we describe the general structure of the approximators. The approximator defined for the DNF-formula of
a function computed at a node $g_i$ in $\beta$ consists of
\begin{enumerate}
\item
  a set $M(g_i) \subseteq {\cal M}_n$ of monomials each of size at most $r$,
\item
  for each monomial $m \in M(g_i)$, a set  $D_i(m) \subseteq \overline{{\cal M}}_n$.
\end{enumerate}
Then, the approximator $\mbox{DNF'}_{\beta}(g_i)$ is defined by
$$
\mbox{DNF'}_{\beta}(g_i) := \bigvee_{m \in M(g_i)} \; \bigvee_{\overline{m} \in D_i(m)} \overline{m}m. 
$$

\smallskip
Consider $i \geq 1$. Assume that the approximators with respect to $g_j$, $j < i$ are already defined. Now we
define the approximator with respect to $g_i$. In dependence of the kind of $g_i$, we distinguish four cases.

\medskip
\noindent
{\em Case 1:\/} $op(g_i) = x_j$, $j \in \{1,2, \ldots,n\}$.

\smallskip
Let
$$
M(g_i) := \{x_j\} \mbox{ and } D_i(x_j) := \{\varepsilon\}.
$$
Then
$$
\mbox{DNF'}_{\beta}(g_i) := \varepsilon x_j.
$$
Obviously, $x_j$ and $\mbox{DNF'}_{\beta}(g_i)$ compute the same function. 

\medskip
\noindent
{\em Case 2:\/} $op(g_i) = \neg x_j$, $j \in \{1,2, \ldots,n\}$.

\smallskip
Let
$$
M(g_i) := \{\varepsilon\} \mbox{ and } D_i(\varepsilon) := \{\neg x_j\}.
$$
Then
$$
\mbox{DNF'}_{\beta}(g_i) := \neg x_j \varepsilon.
$$
Obviously, $\neg x_j$ and $\mbox{DNF'}_{\beta}(g_i)$ compute the same function.

\medskip
\noindent
{\em Case 3:\/} $g_i$ is an $\vee$-gate with direct predecessors $g_{i_1}$ and $g_{i_2}$.

\smallskip
Before the performance of any plucking we have
$$M(g_i) := M(g_{i_1}) \cup M(g_{i_2})$$
and for each $m \in M(g_i)$
$$
D_i(m) := \left\{ \begin{array}{ll}
    D_{i_1}(m) \cup D_{i_2}(m)  & \mbox{if } m \in M(g_{i_1}) \cap M(g_{i_2}), \\
    D_{i_1}(m) & \mbox{if } m \not\in M(g_{i_2}), \\
    D_{i_2}(m) & \mbox{if } m \not\in M(g_{i_1}).
  \end{array}
  \right.
$$
  Now we explain the effect of plucking with respect to these sets. Let $m_1, \ldots,m_p$ be $p$ monomials
  where the corresponding sets $V(m_1), \ldots,V(m_p)$ form a sunflower with core $T$. A plucking
  replaces the monomials $m_1, \ldots,m_p$ by the monomial $m(T)$ consisting of the variables in $T$. Hence, the
  following operations are performed:
$$
  D_i(m(T)):= \left\{ \begin{array}{ll}
    D_i(m(T)) \cup \bigcup_{j=1}^p D_i(m_j) & \mbox{if } m(T) \in M(g_i), \\
    \bigcup_{j=1}^p D_i(m_j) & \mbox{if } m(T) \not\in M(g_i)
  \end{array}
  \right.
$$  
and
$$
M(g_i) := M(g_i) \setminus \{m_1,m_2, \ldots,m_p\} \cup \{m(T)\}.
$$

\medskip
\noindent
{\em Case 4:\/} $g_i$ is an $\wedge$-gate with direct predecessors $g_{i_1}$ and $g_{i_2}$.

\smallskip
Before the performance of any plucking, we have
$$
M(g_i) := \{mm' \mid m \in M(g_{i_1}), m' \in M(g_{i_2}) \mbox{ and $mm'$ has size $\leq r$}\}.
$$

For each $m \in M(g_i)$ for all $m_1 \in M(g_{i_1})$, $m_2 \in M(g_{i_2})$ such that $m = m_1m_2$,
we define
$$
D_i(m,m_1,m_2) := \{\overline{m}\overline{m}' \mid \overline{m} \in D_{i_1}(m_1) \mbox{ and } \overline{m}'
\in D_{i_2}(m_2)\}
$$
and
$$
D_i(m) := \bigcup_{m_1\in M(g_{i_1}), m_2 \in M(g_{i_2}): m_1m_2 = m} D_i(m,m_1,m_2).
$$
Then these sets are modified because of the performed pluckings as described in Case 3.
This finishs the description of the extended approximators.

\medskip
A sunflower-approximator for standard networks consists of two components. One component is the approximated part,
the other component  the non-approximated part. The combinatorial properties of the approximated part are the same as in 
sunflower-approximators for monotone Boolean networks. With respect to the non-approximated part, no combinatorial
properties usable in a lower bound proof can be extracted without any further information about the structure of the
network. Moreover, the non-approximated part causes that arguments used in the lower bound proof for the monotone
complexity do not work now.

The main problem is caused by the performance of pluckings. A necessary property is that the number of inputs in $T_0$
for which a plucking introduces an error can be bounded. Let $m_1,m_2, \ldots,m_p$ be $p$ monomials where the corresponding
sets $V(m_1), \ldots,V(m_p)$ form a sunflower with core $T$. With respect to monotone networks, an input $c \in T_0$ for
which an error is introduced because of the performance of the corresponding plucking cannot fulfill any monomial in
$\{m_1,m_2, \ldots,m_p\}$. Otherwise, the error with respect to $c$ would be already exist before the performance of the
plucking. Exactly this fact is used to get the needed upper bound. But with respect to standard networks, an input
$c \in T_0$ for which an error is introduced because of the performance of the corresponding plucking can fulfill some
monomials in $\{m_1,m_2, \ldots,m_p\}$. This comes from the non-approximated part of the monomials in the approximators.
To see this consider $c \in T_0$ for which an error is introduced. Then there is $\overline{m} \in D_i(m(T))$ such that
$c$ fulfills $\overline{m}m(T)$. By construction, there is $l \in \{1,2, \ldots,p\}$ such that $\overline{m} \in D_i(m_l)$.
Obviously, $c$ cannot fulfill the monomial $m_l$. Otherwise, $c$ would fulfill $\overline{m}m_l$ such that the plucking does
not introduce an error with respect to the input $c$. But with respect to each $j \in \{1,2, \ldots,p\}$ with
$\overline{m} \not\in D_i(m_j)$ $c$ could fulfill the monomial $m_j$ if each monomial in $D_i(m_j)$ contains a literal not
fulfilled by $c$. Therefore, we have to estimate an upper bound for the number of inputs in $c \in T_0$ with
\begin{enumerate}
\item
  $c$ satisfies a monomial $\overline{m}m(T)$ where $\overline{m} \in D_i(m(T))$ and
\item
  for all $1 \leq j \leq p$, $c$ does not fulfill $m_j$ or each monomial in $D_i(m_j)$ contains a literal not fulfilled by $c$.
\end{enumerate}
Without any knowledge about the structure of the monomials in the non-approximated part of the approximators, I see no way to
prove an upper bound for such inputs in $T_0$ which is small enough. This gives evidence that sunflower-approximators alone
cannot be used to prove a super-linear lower bound for the standard complexity of any Boolean function $f \in {\cal B}_n$.

\section{What should be done next?}

To prove a super-linear lower bound for the standard complexity of any explicit Boolean function, we need more
knowledge about the use of negations in a non-monotone Boolean network. Essential for the lower bound proofs for the monotone
complexity of a Boolean function is the following property: In a monotone Boolean network each prime implicant of the function
has to be constructed at the corresponding output node. A standard network computing a Boolean function must not have this
property. Instead of constructing a prime implicant $p$ at the output node, a set $m_1, m_2, \ldots,m_r$ of monomials such that
\begin{enumerate}
\item
  $p$ is a submonomial of each monomial; i.e, $m_i = pm'_i$, $1 \leq i \leq r$ and
\item
  $\bigvee_{i=1}^r m'_i = 1$
\end{enumerate}
could be constructed. To prove a lower bound for the standard complexity of a Boolean function, we have to prove a lower bound
for the number of gates needed for the construction of such a DNF-representation of the function. The problem is that by a
standard network, each algorithm for the computation of a solution could be realized. To clarify this by an example, let us
consider the Boolean function $f$ where the input variables encode an undirected graph on $n$ nodes and $f$ is one iff the
input graph is a $k$-clique; i.e., the graph consists of a $k$-clique and $n-k$ isolated nodes. Note that $f$ is non-monotone and
each prime implicant of $f$ has a literal with respect to each possible edge. For a given graph $G = (V,E)$, it is easy to decide
if $G$ is a $k$-clique. $G$ is a $k$-clique iff $G$ has exactly $k$ nodes with degree $k-1$ and in total $k(k-1)$ edges where each
edge is counted with respect to each end node. $f$ is the {\em exact-clique function}. For checking if exactly $l$ of $m$
variables are one, we can use a non-monotone network of size $O(m)$ \cite[Chapter 3.4]{We}. Therefore, $f$ can be computed by a
standard network of linear size. If we relax the definition of $f$ and we define that $f$ is one iff the input graph $G$ contains
a $k$-clique, the situation changes dramatically. Now $f$ is the NP-complete clique function and it is an open problem if for $f$
a standard network of polynomial size exists. Each deterministic algorithm for the solution of the clique problem can be realized
by a standard network such that the size of the network is polynomial in the time used by the algorithm.

To prove a lower bound for the size of a standard network which computes a given Boolean function $f$, we can only use the
structure of the function $f$. We need an intuition which properties of the prime implicants or the prime clauses of a given
function forces a standard network to use a certain amount of gates. How to get such an intuition? The structure of the prime
implicants of the exact-clique function tells us directly how we can realize the function by a standard network of linear size.
The knowledge about upper bounds may help us to get an
intuition what make the function easy or difficult. Hence, before looking for properties which enable us to prove a lower bound,
I would look for upper bounds for the function under consideration. Which functions should be chosen for the try to get the first
proof of a super-linear lower bound for the non-monotone complexity of an explicit Boolean function?

As for monotone networks, I would first try to obtain a super-linear lower bound for a Boolean function with many outputs
as the Boolean convolution, the Boolean matrix multiplication or (1,1)-disjoint Boolean sums. A non-constant lower bound with
respect to each output would result in a super-linear lower bound for the function. Firstly, upper bounds for the
chosen function should be investigated to learn something about the usefulness of negations with respect to the considered
function. Can we adapt any method developed for monotone networks to obtain a super-linear lower bound? The paper of 
Wei{\ss} \cite{Wei} could be a good starting point.

\medskip
How to proceed the work with respect to the P versus NP problem? Currently, I am convinced that we are far away to prove a
super-polynomial lower bound for the non-monotone complexity of any explicit Boolean function. On the other hand, the
strongest barrier towards proving $\mbox{P} \not= \mbox{NP}$ could be that it holds $\mbox{P} = \mbox{NP}$. To ensure that
the whole time spent for working on the P versus NP problem is not used to prove an impossible theorem, I would switch to the
try to develop a polynomial algorithm for the solution of an NP-complete problem. Moreover, also in the case that
$\mbox{P} \not= \mbox{NP}$, understanding why it is not possible to develop a polynomial algorithm for the solution of the
considered NP-complete problem could be of help to prove a lower bound for the standard complexity of the corresponding
Boolean function. What kind of NP-complete problem should be chosen? I think that a good candidate would be an NP-complete
optimization problem for the following reasons.

A general approach for the solution of an optimization problem is the following: Start with a feasible solution of the
optimization problem under consideration. As long as possible apply to the current feasible solution an improvement step. The
improvement step replaces a part of the current solution by a part which is outside of the current solution such that the
obtained solution is feasible and the value of the objective function is improved. To get a polynomial algorithm for the
optimization problem, the following properties should be fulfilled:
\begin{enumerate}
\item
  A suboptimal feasible solution always allows the application of an improvement step,
\item
  after a polynomial number of improvement steps, an obtimal solution is obtained, and
\item
  an improvement step can be performed in polynomial time.
\end{enumerate}
A well known optimization problem where this approach has led to a polynomial time algorithm is the maximum matching problem.
Let $G = (V,E)$ be an undirected graph. $M \subseteq E$ is a {\em matching} if no two edges in $M$ have a common node. A
matching $M$ is {\em maximal} if there is no edge $e \in E \setminus M$ such that $M \cup \{e\}$ is a matching. A matching $M$
is {\em maximum} if there exists no matching $M' \subseteq E$ of larger size.  A maximal matching $M$ is {\em minimum} if there
is no maximal matching $M'$ of $G$ such that $|M'| < |M|$. Given an undirected graph $G = (V,E)$, the
{\em maximum matching problem} is finding a maximum matching $M \subseteq E$. The {\em minimum maximal matching problem} is
finding a minimum maximal matching $M \subseteq E$. Note that the minimum maximal matching problem is NP-complete \cite{GaJo}.
Let $M \subseteq E$ be a matching of $G$. A node $v \in V$ is {\em $M$-free} iff $v$ is not incident to an edge in $M$.

In 1891, Peterson \cite{Pe} introduced the technique of alternating paths. A path $P = v_0,v_1,\ldots,v_k$ is {\em $M$-alternating}
if it contains alternately edges in $M$ and in $E \setminus M$. Let $P = v_0,v_1, \ldots,v_k$ be a simple $M$-alternating path.
$P$ is {\em $M$-augmenting} if $v_0$ and $v_k$ are $M$-free. $M \oplus P$ denotes the {\em symmetric diffence\/} of $M$ and $P$;
i.e., $M \oplus P = M \setminus P \cup P \setminus M$. If $P$ is an $M$-augmenting path then $M \oplus P$ is a matching of $G$, and
$|M \oplus P| = |M| + 1$. In 1957, Berge \cite{Be} proved that a matching $M \subseteq E$ is maximum iff there is no $M$-augmenting
path in $G$. Until 1963, for non-bipartite graphs only exponential time algorithms for the construction of $M$-augmenting paths
has been known. Then in 1963, Edmonds \cite{Ed} has shown how to construct an $M$-augmenting path in a non-bipartite graph in
polynomial time if an $M$-augmenting path exists. This resulted in a polynomial algorithm for the maximum matching problem.

Berge's characterization theorem has resulted in the construction of an improvement step. Edmonds has shown how this improvement
step can be performed in polynomial time. If we try to apply an analogous approach to an NP-complete optimization problem, we need
such a problem which allows the proof of a characterization theorem which can be used for the construction of an improvement step.
Then we can try to develop a polynomial implementation of the improvement step. I think that the minimum maximal matching problem
could be a good candidate for such an NP-complete problem.

\section*{Acknowledgment}

    I would like to thank Stasys Jukna for many helpful discussions in 2017 after my mistake.

\end{document}